\documentclass[letterpaper,twoside,12pt]{article}
\usepackage{amssymb}
\usepackage{amsmath}
\usepackage{latexsym}
\usepackage{verbatim}
\usepackage{graphicx}
\usepackage{epsfig}
\usepackage{stmaryrd}
\usepackage{rotating,graphicx}
\usepackage[english]{babel}
\usepackage{setspace}
\usepackage[english]{babel}
\usepackage[utf8]{inputenc}
\usepackage[T1]{fontenc}
\usepackage{lmodern}

\begin{document}

\newcommand{\bra}[1]{\langle #1|}
\newcommand{\ket}[1]{|#1\rangle}
\newcommand{\braket}[2]{\langle #1|#2\rangle}

\begin{Large}
\begin{center}
\textbf{Measurements according to Consistent Histories}\\
\end{center}
\end{Large}

\begin{center}
\begin{large}
Elias Okon\\
\end{large}
\textit{Instituto de Investigaciones Filos\'oficas, Universidad Nacional Aut\'onoma de M\'exico, Mexico City, Mexico.\\ E-mail:} \texttt{eokon@filosoficas.unam.mx}\\[.5cm]
\begin{large}
Daniel Sudarsky\\
\end{large}
\textit{Instituto de Ciencias Nucleares, Universidad Nacional Aut\'onoma de M\'exico, Mexico City, Mexico.\\ E-mail:} \texttt{sudarsky@nucleares.unam.mx}\\[.5cm]
\end{center}

We critically evaluate the treatment of the notion of measurement in the Consistent Histories approach to quantum mechanics. We find such a treatment unsatisfactory because it relies, often implicitly, on elements external to those provided by the formalism. In particular, we note that, in order for the formalism to be informative when dealing with measurement scenarios, one needs to assume that the appropriate choice of framework is such that apparatuses are always in states of well defined pointer positions after measurements. The problem is that there is nothing in the formalism to justify this assumption. We conclude that the Consistent Histories approach, contrary to what is claimed by its proponents,  fails to provide a truly satisfactory resolution for the measurement problem in quantum  theory. 



\section{Introduction}
\onehalfspacing
The Consistent Histories (CH) approach to quantum mechanics (also known as Decoherent Histories or Consistent Quantum Theory) is an interpretation of quantum theory that, according to its proponents, overcomes the measurement problem. The formalism is supposed to assign probabilities to histories of all kinds of systems, microscopic or macroscopic, using a single and universal machinery,  and without any reference to measurements. As a result, CH is claimed  to  require  neither a separation of the situation under consideration into classical and quantum parts, nor a special treatment for  the measurement situations, thus solving the measurement problem. In this work, however, we challenge such assertions. To do so, we evaluate in detail its  treatment of the notion of measurement. We find such treatment unsatisfactory because it relies, often implicitly, on elements external to those provided by the formalism. Furthermore, we argue that the introduction of these external notions invalidates the claim that CH solves the measurement problem. We conclude, then, that the CH approach fails to provide a truly satisfactory solution to the measurement problem of quantum mechanics.

Our manuscript is organized as follows: in section \ref{MPCH}, we introduce the measurement problem afflicting the standard formulation of quantum mechanics and  we describe the CH formalism. In section \ref{MCQT}, we evaluate the treatment of measurements in CH and we present our main arguments against the claim that CH solves the measurement problem; we also discuss possible objections to our arguments. In section \ref{Alt}, we comment on a couple of possible replies of a more general scope against our criticism, regarding the necessity for, 
and  the existence of alternatives to CH, and in section \ref{C}, we present our conclusions.
\section{The Measurement Problem and the Consistent Histories Approach}
\label{MPCH}
The measurement problem, as commonly understood, is the fact that, even though the standard formulation of quantum mechanics depends crucially on the notion of measurement (to decide when to use the unitary evolution and when the reduction postulate), such notion is never formally defined within the theory. Then, in order to apply the formalism, one needs to know, by means external to quantum mechanics, what constitutes a measurement, when a measurement is taking place, and what it is  one is measuring. 

An alternative way to present the problem is as a mismatch between experience and some predictions of \emph{unitary} quantum mechanics. More concretely, the problem corresponds to a discrepancy between the prediction of the widespread presence of macroscopic superpositions (so-called Schr\"odinger cat states), and the fact that observers always end up with determinate measurement results. 

A more formal statement of the measurement problem asserts the mutual incompatibility of the following three statements (see \cite{Mau:95}): i) the wave function of a system is complete, ii) the wave function always evolves according to  a linear dynamical equation and iii) measurements always have determinate outcomes. This is a particularly useful way to state the problem in order to classify the  attempts to resolve it. So, for example, hidden variable theories negate i), objective collapse models negate ii) and many-worlds scenarios negate iii).\footnote{As we will see below, the wave function does not play a central role in CH so it could be argued that the formalism does not satisfy i) and ii). As for iii),  CH satisfies it \emph{only if the right framework is chosen}.}   

Bohr's famous response to the measurement problem involves taking a purely operational stance. That is, it involves holding that quantum mechanics is a theory that only talks about probabilities for outcomes of measurements performed by observers, and nothing more. The problem is that such a position is untenable if one wants to think about the theory as the basis for a complete description of a physical world with ontological independence from us and which lays beyond ourselves and our brains.

 Another popular attempt to address the question of when  to rely on  the   evolution as provided  by  the  Schr\"odinger equation is to hold that, for a given micro system S, that equation should be used  at all times  except at those when the macro-world of measuring devices and observers interacts with S. This attempt is, of course, unsatisfactory if one believes that quantum theory should apply to everything, and that there is no fundamental wall separating the macro-world from the micro-world. Clearly, the issue becomes even more serious if one wants to apply quantum mechanics to cosmology (see \cite{Bel:81}). 

A possible objection against the claim that the standard formulation of quantum mechanics suffers from the measurement problem is to argue that, in any theory, some pre-theoretical notions are required in order to make sense of measurements; that some  questions, such as what part of the system is to be considered a `macroscopic apparatus' are also not answered in other theories, like classical mechanics, where one does not think that there is a measurement problem. The key point, however, is that, in other theories, such ability to determine what part of the system is to be considered a `macroscopic apparatus' is \emph{not} required in order to use the theory to make predictions. In standard quantum mechanics, in contrast, the need to separate the system from the measuring apparatus is essential in order to extract predictions from the formalism. Therefore, from this point of view, a solution to the measurement problem would be given by a formalism capable of making predictions without requiring the introduction of external notions of  measurements or of anything else (in the next section we will explain in more detail what constitutes a satisfactory solution to the measurement problem).


The CH approach to quantum mechanics is regarded by its proponents as an interpretation of quantum theory that overcomes the measurement problem. It was first introduced in 1984 by Griffiths in \cite{Gri:84}, and developed in the subsequent years by himself, Omnès, Gell–Mann and Hartle in \cite{Gri:86,Gri:87,Omn:87,Omn:88,Gel.Har:90,Gel.Har:93}. Contemporary presentations of the formalism include \cite{Gri:03,Gri:09,Hoh:10,Gri:11,Gri:13}, ref. \cite{Gri:03} being regarded as the most detailed and complete discussion of the conceptual foundations of CH available. 

In fact, CH recognizes not one but two different measurement problems (see \cite{Gri:13}). The first one is precisely the one we have been talking about, namely, the fact that if a measurement apparatus is treated quantum mechanically, then the Schrödinger evolution will typically drive it into a superposition of macroscopic states and no definite outcomes will be obtained. In \cite[p. 214]{Gri:03}, for example, the orthodox solution to the measurement problem is criticized as follows:
\begin{quotation}
\noindent It seems somewhat arbitrary to abandon the state... obtained by unitary time evolution,... without providing some better reason than the fact that a measurement occurs; after all, what is special about a quantum measurement? All real measurement apparatus is constructed out of aggregates of particles to which the laws of quantum mechanics apply, so the apparatus ought to be described by those laws, and not used to provide an excuse for their breakdown.
\end{quotation}
The second measurement problem has to do with the impossibility, within the standard theory, of relating the pointer positions of a measurement apparatus with the state of the measured microscopic system \emph{before}  a measurement takes place. That is, with the fact that, in standard quantum mechanics, measurements typically disturb the measured system and, as a result, the values of the measured properties before and after the measurement are not related.

According to its proponents, CH constitutes a \emph{realist} approach to quantum theory that solves the measurement problem (actually both of them), as well as a whole range of quantum paradoxes. The claim is that the formalism assigns probabilities for all systems, microscopic or macroscopic, using the same machinery and without any reference to measurements. Therefore, actual laboratory experiments can be analyzed in purely quantum terms, using the same principles that govern any quantum system. Consequently, CH is held not to rely on the notion of measurement, and not to require an artificial separation between the classical and quantum domains. For instance, in \cite[p. 3]{Har:93} one reads: ``In this formulation, it is the internal consistency of probability sum rules that determines the sets of alternatives of the closed system for which probabilities are predicted rather than any external notion of `measurement'.''

In broad terms, the formalism that (allegedly) accomplishes the above, works as follows. One starts with the notion of a \emph{history}, which is a sequence of properties of a quantum system. Such properties are represented by projection operators on the Hilbert space at successive times. Then, one introduces sets of histories and specifies rules that assign probabilities to the various elements of each of  the sets. However, not all sets or families of histories allow for probabilities to be assigned. This assignment can be done provided that the sum of probabilities of the histories in the family equals one, and that all pairs of histories within the family are orthogonal (in the sense that there is no overlap between them). Families satisfying these two conditions are called \emph{frameworks}, or \emph{realms}, and represent the only sets of histories for which probabilities can be consistently defined. Frameworks, then, constitute the quantum counterpart of sample spaces in ordinary probability theory. It is important to note that, according to CH, within every framework, every observable compatible with this framework has a definite value.\footnote{This is clear within Griffith's formulation of CH but it is not so straightforward in Gell–Mann and Hartle's formulation (see \cite{Oko.Sud:14a}).}  Furthermore, the quantum state does not determine these definite values (not even within a suitable framework), but only gives probabilities. This is because, according to CH, these definite values change indeterministically (see \cite{Gri:11}).

There is, however, an apparent complication: given a quantum system, multiple incompatible frameworks can be constructed. In order to avoid inconsistencies, the following rules or principles must be enforced (see \cite[p. 98]{Gri:13}):
\begin{itemize}
\item \textbf{Single-Framework Rule}: probabilistic reasoning is invalid unless it is carried out using a single framework. 
\end{itemize}
Therefore, any prediction of the theory must be constrained to a single framework, and statements belonging to incompatible frameworks cannot be compared in any way.
\begin{itemize}
\item \textbf{Principle of Liberty}: one can use whatever framework one chooses in order to describe a system.
\item \textbf{Principle of Equality}: all frameworks are equally acceptable in terms of fundamental quantum mechanics.
\item \textbf{Principle of Utility}: not all frameworks are equally useful in answering particular questions of physical interest.
\end{itemize}
As a result, the CH formalism violates the following principle (see \cite[p. 99]{Gri:13}):
\begin{itemize}
\item \textbf{Principle of Unicity}: alternative descriptions of physical systems always can be combined into a single unified description from which all views can be derived as partial descriptions.
\end{itemize}
The reason for this is that, according to the theory, no single framework suffices to characterize a quantum system completely. Therefore, quantum systems in particular, and reality in general, are characterized by various alternative and incompatible descriptions.

As we remarked above, and as can be noted from our description of the formalism, \emph{measurements} play no special role within CH. Standard, laboratory measurements are then just another type of physical process which can, as any other process, be described within the theory. In order to do so, an appropriate framework must be selected within which probabilities, identical to those given by applying Born's rule, can be calculated. Therefore, even though measurements play no fundamental role within the theory, the CH formalism appears to be capable of describing them consistently, yielding the same results as the orthodox interpretation. 

All of the above sounds very appealing; the CH formalism seems to avoid the notion of measurement and to give rules that are unified in the treatment of micro and macro systems. Now, the question is whether the reference to measurement as a special and independent notion, which was so bothersome within the Copenhagen interpretation, has really been removed from the formalism. Or, more concretely, if the CH formalism really deals with measurement situations with exactly the same tools that it does in other cases. We believe it does not. That is because, in  contrast to what CH proponents claim,   it is not the case that, when  considering  measurement situations,  all  the frameworks are equally valid (principle of Equality) and that  it  is  just  that one is more useful or informative than the others  (principle of Utility). The situation, instead, is that there exists only one framework that correctly describes what in fact we perceive or experience. And the problem we want to stress is that CH, as standard quantum mechanics, is incapable of explaining and predicting, without relying on elements external to the theory (such as our intuition or experience regarding outcomes of experiments), which one is the framework that will correspond to our perceptions.

We claim, then, that the special status given to measurements remains, but that it is hidden in the unspoken rules that indicate which framework one should use when considering any specific measurement situation. But what are these unspoken rules? As we will show below, applications of the formalism to actual experiments, as discussed, for example, in \cite{Gri:03,Hoh:10,Gri:13}, require the incorporation of elements that are external to the CH framework. Furthermore, these external notions, in essence, bring back, in a hidden way, some basic aspects of the Copenhagen rules for the treatment of measurements (e.g., the notion that after a measurement, a measuring apparatus is always in a state of well defined pointer position). Therefore, when discussing measurement situations, there is a reliance on an implicit rule, often presented  as ``common sense,''  whereby the appropriate framework to be used is such that macroscopic apparatuses always have well defined pointer positions. 
The problem, of course,  is that there is nothing in the formalism to justify this assumption.

We hold, therefore, that contrary to what is claimed, the formalism offered simply fails to resolve the measurement problem (in fact, it fails to resolve both of the measurement problems described in \cite{Gri:13}). It seems, actually, that in this respect CH does not even improve upon the Copenhagen interpretation. Furthermore, given that, after all, measurements are our only way to verify the validity of any theory, the inability of the CH formalism to deal with them represents a serious complication. In the next section we explain in more detail what we find unsatisfactory about the treatment of measurements in CH.

\section{Measurements according to Consistent Histories}
\label{MCQT}
As we mentioned above, the measurement problem can be stated as the fact that the notion of measurement is essential, but never formally defined, within standard quantum mechanics. Therefore, in order to apply the standard formalism, one needs to know, by means external to quantum mechanics, what constitutes a measurement, when a measurement is taking place, and what it is that one is measuring. A solution to the measurement problem, then, would be a version of quantum theory that can be used, i.e., that is able to make predictions, without requiring the introduction of elements external to the theory. What we expect from such a theory is to inform us on the possible outcomes of experiments, along with their corresponding probabilities, and not only to provide us with probabilities for a list of outcomes that must be provided by considerations external to the theory (as is the case, for example, in the Copenhagen interpretation, where one needs to know \emph{a priori} what constitutes a measurement, what it is that a particular apparatus measures, etc.).

Examples of theories that accomplish the above include objective collapse models (such as the one developed by Ghirardi, Rimini and Weber (GRW) in \cite{GRW:86}  or the Continuous Spontaneous Localization (CSL) theory of Pearle, \cite{Pea:89}), and Bohmian Mechanics \cite{Bohm}. In these theories, a complete specification of the initial state of a system is enough, without the need of any other input, in order to predict the possible time evolutions of the system, along with their corresponding probabilities (something that, as we will see, cannot be said about CH). It is worth noting that it is not uncommon to encounter proposals that claim to address the measurement problem, but that implicitly assume that the measuring apparatuses are always found in definite positions for their pointer degrees of freedom. That is, they assume a big part of what needs to be explained.

In this section, we will critically examine applications of the CH formalism to \emph{concrete} measurement situations, particularly as developed in \cite{Gri:03,Hoh:10,Gri:13}. Our main objections will stem from the fact that, on such applications, and contrary to what one would expect from CH, no use is actually made of the Principles of Equality, Liberty or Utility. Instead, the procedure employed relies on elements external to the explicit CH formalism. Moreover, such external input is nowhere systematized but depends, implicitly, on the use of ``common sense.'' More precisely, we believe that when applying CH to a concrete measurement situation, it is not the case  that all frameworks are equally valid, that one has the liberty to use whichever framework one decides and that it is just that one is more useful or informative than the others. The situation, instead, is that among all the possible frameworks, only one is suitable to describe what in fact we perceive or experience. And, as we said, the main problem is that CH is incapable of recognizing in advance, and without bringing in elements that rely on our intuition and experience, which is going to be the framework that does the job. We conclude, then, that CH does not really constitute a satisfactory solution to the measurement problem. 

In order to support our claims, the plan for this section is as follows. We will begin by making explicit the implicit dependence on external input of applications of CH to measurement situations. To do so, we will examine in detail concrete examples of such applications presented in the above mentioned texts. After that, we will consider and criticize two popular claims regarding framework selection for measurement situations, often given as replies to the kind of objections we present. We  are referring  to the  arguments   that frameworks must be chosen i) in order to model the experimental situation at hand or ii) according to the questions one is interested in answering. We will show that such proposals do not improve the situation regarding the need for external input. To conclude this section, we will make a few comments regarding CH's alleged solution to the so-called second measurement problem.

So, why do we claim that the application of CH to measurement situations requires external input? First of all, in the description of the procedure presented in, for example, \cite{Gri:03}, there are statements that indicate, rather explicitly, the dependence of the formalism on extraneous elements (without any explicit reference to CH principles). For instance, in discussing the Stern-Gerlach experiment in section 17.2, the book states (all emphasis in bold is ours):
 \begin{quotation}
\noindent The unitary history... cannot be used to describe the measuring process, because the measurement outcomes... are \textbf{clearly incompatible} with the final state [of the unitary history]... A quantum mechanical description of a measurement with particular outcomes must, \textbf{obviously}, employ a framework in which these outcomes are represented by \textbf{appropriate projectors}... (\cite[p. 202]{Gri:03})
 \end{quotation}
The  issue is then,  how do we know which  ones  are these ``appropriate projectors''? It seems that, to do so, one must know, by means \emph{external} to the formalism, what are the possible outcomes of the experiment. It is then necessary to adapt the selection of the framework according to that knowledge. 
The application of the formalism, then, relies implicitly on the practical fact that in laboratory situations we know what is being measured, which part of the system is to be considered an apparatus, and what are the possible results. In other words, we must use our experience with macroscopic apparatuses as part of the input for the analysis. That means that the theory does not seem to account for those experiences but needs them as extra input, which seems to defeat the purpose of going beyond the Copenhagen interpretation. 

There are various other examples of such kind of statements appearing in the analysis presented in \cite{Gri:03}. For instance:
\begin{quotation}
\noindent It is customary to use the term \emph{pointer basis} for an orthonormal basis, or more generally a decomposition of the identity such as employed in the generalized Born rule... that allows one to discuss the outcomes of a measurement \textbf{in a sensible way}. (\cite[p. 115]{Gri:03})
  \end{quotation}
The problem, again, is that this ``sensible way'' of choosing the framework constitutes an element extraneous to the theory. It represents information not present in the formalism itself, crucially required in order for the theory to deliver an accurate description of what in fact we observe in experiments. In other words, the formalism is unable to inform us about the possible outcomes of a given measurement;  this information must be  put in  by hand while choosing a ``sensible'' basis.

A common response of CH proponents to the type of objections raised above is that the selection of a framework for measurement situations is not put in by hand but that it is chosen in order to ``model the particular experimental situation designed by the experimentalist.'' Accordingly, one reads
\begin{quotation}
\noindent There is, to be sure, an alternative unitary framework in which the fearsome cat... is present..., and physicists, philosophers and science fiction writers are at Liberty to contemplate it as long as they keep in mind its \textbf{incompatibility with} (and thus irrelevance to) \textbf{the sorts of descriptions commonly employed by competent experimental physicists when describing work carried out in their laboratories}. (\cite[p. 105]{Gri:13})
  \end{quotation}
or even more explicitly,
\begin{quotation}
\noindent At the time... just before the measurement $|\psi_0\rangle$ is incompatible with the... [projective decomposition] corresponding to \textbf{the properties... the apparatus has been designed to measure}, a framework in which $|\psi_0\rangle$  makes sense will not be useful for discussing the measurement \emph{as a measurement}, i.e., as measuring \emph{something}, so the physicist interested in that aspect of things must use something else. (\cite[p. 105]{Gri:13})
  \end{quotation}
However, the fact that a given measuring apparatus \emph{actually} measures some property is something that cannot be deduced from the CH formalism but that must be discovered by experience. That is, 
if one is given a new measurement equipment, described entirely in quantum terms (via a Hamiltonian, an initial state, etc.), CH is unable to answer, unlike, say, CSL or Bohmian mechanics, which are going to be its possible final states when we use it in the laboratory. Therefore, in a given situation, described by providing the complete physical set-up in terms of the initial state of the closed system and the Hamiltonian, CH is incapable of predicting which framework one must choose.
Another typical response, regarding framework selection for measurements, (allegedly) relies directly on the Principle of Utility.  The idea is that the choice must be made ``according to the questions one is interested in answering.''
Therefore, one finds
\begin{quotation}
\noindent  A measurement of $S_x$, for example, needs to be interpreted in the $x$ framework. It will not be informative if analyzed in the $z$ framework. (\cite[p. 2838]{Hoh:10})
  \end{quotation}
or, regarding unitary evolution
\begin{quotation}
\noindent This is a valid prediction of quantum mechanics, but unless the final state is the eigenstate of some \textbf{straightforward observable} it is also not a particularly useful prediction since it contains no accessible physical information. (\cite[p. 2841]{Hoh:10})
  \end{quotation}
or
\begin{quotation}
\noindent It is true, of course, that in the unitary framework... the wave function constitutes a valid history, but that history is, in general, \textbf{not useful for answering physical questions} about the system... (\cite[p. 2843]{Hoh:10})
  \end{quotation}
The trouble that we have with this utilitary answer, to begin with, is that it seems to hide the fact that, for a given experimental set-up, there is \emph{only one} framework that will yield probabilities that match the actual results. As a consequence, for measurement situations, the Principles of Equality and Liberty do not seem to apply. Of course, supporters of CH could argue that the Principle of Utility is totally compatible with the fact that there is only one framework that correctly describes the actual measurement outcome, and that, while the Principles of Equality and Liberty are of no use for describing measurements and their outcomes, they still hold in such situations. At any rate, the truly problematic point, again, is that the CH formalism is incapable of picking out the right framework. That is, it does not offer any clear characterization regarding the exact relation between the experimental set-up and the framework one needs to use in order to get reliable predictions out of the theory. The most detailed response to our objection appears in \cite[p. 2838]{Hoh:10} where, when discussing the measurement of a spin, it is stated that
\begin{quotation}
\noindent In the $x$ framework any state has two disjoint possibilities or properties $[x^+]$ and $[x^-]$, so for a system prepared in the state $|z^+\rangle$ a measurement of $S_x$, when viewed in the $x$ framework, will reveal one or the other of these properties, each with probability $\frac{1}{2}$. In the $z$ framework, on the other hand, I am unable to interpret an $S_x$ measurement.
  \end{quotation}
However, it remains unclear what is supposed to be the relation between the framework we decide to use and what we in fact observe when we perform the experiment. So, for example, what would change in what we \emph{in fact observe} if we decide to use the $z$ framework when performing a measurement of $S_x$? Clearly nothing would. It seems then that the choice of framework must be done, not according to the questions one is interested in answering but according to the actual experimental set-up. The problem, once more, is that CH is unable, given a particular experimental set-up described in quantum terms, to indicate which, among all the possible ones, is the appropriate framework. Furthermore, even according to \cite{Gri:03}, for a given experimental set-up there is, among all the sets offered by the formalism, only one that correctly describes the outcomes. If this is the case, then it seems to us, not only that the formalism is incomplete, but also that the other frameworks play no role whatsoever in the application of the formalism to particular situations.

To close this section we would like to make one last comment about what in \cite{Gri:13} is referred to as the second measurement problem, which is the fact that the standard theory is incapable of relating the pointer positions of a measurement apparatus with the state of the measured microscopic system \emph{before} measurements take place. CH is supposed to be able to solve the problem because, for example, in a EPR setting, one can meaningfully talk about ``particles possessing definite spin in  advance of measurements.'' So for example, when discussing EPR, \cite{Gri:03} explains
\begin{quotation}
\noindent Even stronger results can be obtained using the family... in which the stochastic split takes place at an earlier time. In this family it is possible to view the measurement of $S_{az}$ as revealing a pre-existing property of particle $a$ at a time before the measurement took place, a value which was already the opposite of $S_{bz}$. (\cite[p. 282]{Gri:03})
\end{quotation}
It seems then that one can say that some particle had a definite property before a measurement took place only if one chooses a specific framework where this is so. But, if according to CH, all frameworks are equally valid, it is not clear why the description according to this one framework should be taken that seriously. That is, the problem is solved in only one of an infinite number of possible frameworks.\footnote{See \cite{Esp:87} for a stronger argument against the claim that CH solves what \cite{Gri:13} calls the second measurement problem.} 


\section{Alternatives}
\label{Alt}
Before wrapping up, we would like to comment on a couple of possible broad scope replies against our criticism having to do with the necessity for, and the existence of alternatives to CH. The first one holds that the CH formalism is capable of making predictions for all experiments or observations suitably specified, and that that is enough for physics, regardless of the conceptual issues we might have problems understanding. The second maintains that it would be better if we could propose instead an alternative formulation of quantum mechanics of comparable scope and clarity that has the features we seem to require.

Regarding the first point, we do not believe it is compelling because, as we explained in the previous section, without the aid of external input, the CH formalism is incapable of making specific predictions. As we showed above, in order for the formalism to do so for a given experiment, one must put in by hand what are the possible outcomes for the experiment, a fact that one knows from experience, but cannot be predicted from the formalism. Without such external information, the formalism is unable to correctly describe what we in fact observe in the laboratory. Another way to say this is that in order to make predictions, one must specify the observables to be measured and the times of such measurements. The problem is that at such point the measurement problem resurfaces because we need to know what constitutes a measurement, something not specified within CH. On the other hand, if one wants to simply specify the physical systems and its Hamiltonian (including within the system the quantum description of the measuring devices and the interactions of those and the system), then no predictions emerge.

To illustrate this, consider again the Stern-Gerlach example of sec. 17.2 of \cite{Gri:03} quoted above. Without the external information about what are the correct measurement outcomes it is not possible to argue that some framework, containing a given final state, is not acceptable. Therefore, without that extraneous input, one cannot select the framework, and without it one cannot assign probabilities. Regarding the fact that CH makes predictions equivalent to those given by the Born rule, we agree that that is the case, but only after the external information about the possible final states for the measuring apparatuses is introduced. However, we do not find that surprising because, by introducing such external input, the standard Born rule itself is being inadvertently introduced.

What about the availability of an alternative formulation of comparable scope and clarity to CH but with the features that we require? First of all, it is important to make clear that the assessment of CH as a  solution to the measurement problem is independent from the existence of viable alternatives. That is, the lack of an alternative in no way would cancel the fact that CH does not solve the problem. Having said that, do we think we already have a fully satisfactory version (or modification) of quantum theory that is free of inconstancies and agrees with observations? The short answer is: no. However, let us give a more appropriate response by saying the following:
\begin{enumerate}
\item We are quite convinced that, due to the measurement problem, the standard interpretation is not satisfactory (and that CH does not offer a solution).
\item We do not claim at this point to have a fully satisfactory alternative.
 \item We believe that both Bohmian mechanics \cite{Bohm} and objective collapse models (along the lines of GRW \cite{GRW:86} or CSL \cite{Pea:89}) do offer promising paths. We might,  of course, be proven wrong here.
\end{enumerate} 
Moreover, we believe that it is possible to empirically test objective collapse theories by combining them with models of inflationary cosmology and by comparing the predictions to the precise data from the Cosmic Microwave Background. It is also possible to set bounds on the parameters of such models with the aid of laboratory experiments, \cite{BLS}. However, as such an enterprise involves a radical idea (i.e., modifying quantum theory), it is important to make sure that something like that is really needed. That is, we must be sure that there is a problem that the existing proposals fail to resolve in a fully satisfactory way. In this sense, we do see the CH approach, as well as the proposals of interpretational adjustments based on decoherence, as some of the most powerful attempts to avoid the conclusion that quantum theory needs to be modified. Unfortunately, we do not believe they succeed.

\section{Conclusions}
\label{C}

In this manuscript, we have critically analyzed a key element of the CH proposal to solve the measurement problem of quantum theory, namely its treatment of the notion of measurement (a notion that CH does not take to be fundamental).

We have started by identifying the essential difficulty involved in the measurement problem, and by stating what would constitute a satisfactory solution. In this regard, we have stressed the need to provide a unified treatment for all physical systems where no privileged status is  given to any subsystem simply because it is identified as an observer or as a measuring device. Therefore, the treatment of such subsystems should not involve special and distinct rules, over and above those which are supposed to rule the quantum mechanical characterization of the behavior of any other subsystem.
  
We have seen that the CH formalism fails to provide a truly satisfactory resolution for the problem precisely because, contrary to what is claimed by its proponents, when treating measurements it cannot avoid relying, often in an implicit way, and other times by disguising them as ``common sense,'' on elements external to the formalism. We claim, as a result, that CH does in no way succeed in liberating the notion of measurement from being exceptional. Furthermore, we have pointed out that, contrary to what CH proponents hold, for measurement set-ups, it is not the case that all frameworks are equally valid, and that it is just that some of them  turn out to be more useful to analyze the situation than others. Instead, there is a \emph{particular} framework with which it is possible to make predictions that agree with the actual results. The problem, though, is that the CH formalism cannot, without external \emph{unsystematized} input, single out such privileged framework. Moreover, as we have seen, these external notions rely implicitly on the basics of the Copenhagen interpretation.  We conclude, then, that regarding situations involving measurements, which, at the end of the day, are the only ones that are relevant in order to evaluate the empirical content of a theory, CH does not improve upon standard quantum mechanics. And, given that that was its expressed purpose, we can safely claim that it fails in its endeavor.
 
We hope that this work contributes to research on the foundations of quantum mechanics by helping to focus attention on the deficiencies of what seems, at first sight, to be a very promising path toward the solution of the measurement problem. Moreover, we hope that this kind of study will encourage others to search for alternative ideas that might lead us to the proverbial light at the end of the tunnel.    

\section*{Acknowledgments}
We wish to thank Robert Griffiths and James Hartle for very helpful e-mail discussions; we also thank an anonymous referee for urging us to sharpen some of our arguments and for helping us improve the overall presentation of the manuscript. We acknowledge partial financial support from DGAPA-UNAM projects IN107412 (DS), IA400114 (EO), and CONACyT projects 101712 and 220738  (DS).
\bibliographystyle{plain}
\bibliography{biblioCH.bib}
\end{document}